\documentclass[aps,pra,twocolumn,showpacs,groupedaddress]{revtex4-1}

\usepackage{graphicx}  
\usepackage{dcolumn}   
\usepackage{bm}        
\usepackage{amssymb}   
\usepackage{amsmath}
\usepackage{braket}
\usepackage[mathscr]{euscript}
\usepackage{color}
\usepackage{epsfig,color}
\usepackage{hyperref}

\usepackage{wrapfig}
\usepackage{lipsum}
\usepackage{graphicx,kantlipsum,setspace}
\usepackage{caption}
\usepackage[labelfont=bf,labelsep=quad,justification=justified]{caption}

\begin{document}
\title{Non-Hermitian Optics}
\author{Amarendra K Sarma}
\email{aksarma@iitg.ernet.in}
\affiliation{Department of Physics, Indian Institute of Technology Guwahati\\ Guwahati-781039, Assam, India}
\date{\today}
\begin{abstract}
Non-Hermitian optics, mostly known as {\it Parity-time symmetric optics}, is considered as one of the frontiers areas of research in optical sciences at present. This area is largely inspired by the so-called non-Hermitian quantum physics. While the non-Hermitian quantum mechanics is yet to be accepted widely, parity-time symmetric optics is already a craze among physicists with many experimental results and demonstrations to its support. In this article, keeping under graduate students in mind, in particular, we are giving a brief introduction to this promising area of research in physics.
\end{abstract}
\maketitle
\section{\label{sec1}Introduction}

In a first course on Quantum Mechanics, we are taught that the Hamiltonian, $H$, describing a quantum system needs to be hermitian, i.e. $H=H^{\dagger}$. This is because, the eigenvalues corresponding to a hermitian Hamiltonian is always real; and eigenvalues are directly related to physically observable quantities in experiments. While the quantum mechanics we are familiar with is still quite successful, in 1998 Carl Bender and S. Boettcher \cite{1} put forward a completely new perspective. They said that even non-Hermitian Hamiltonians could exhibit real eigen-spectra provided they obey the parity-time symmetry, resulting in the so-called parity-time symmetric quantum mechanics. Before we understand what is meant by parity-time symmetry, it should be kept in mind that this seemingly new quantum mechanics is not in conflict with the usual quantum mechanics, rather it is an extension of the conventional quantum mechanics into the complex domain.
In fact, the story began by the close investigation of the following class of $\mathcal{PT}$-symmetric Hamiltonians, having the form:

\begin{align}
H=p^2+x^2(ix)^{\epsilon}
\end{align}  
where $p$ and $x$ are momentum and position respectively, while ${\epsilon}$ is a real parameter and $i=\sqrt{-1}$. The Hamiltonians,$H$, are $\mathcal{PT}$symmetric because under space reflection, $x$ changes sign while under time reversal $i$ changes sign. Hence under the combined operation of parity and time reversal, the  Hamiltonian remains invariant. To put mathematically, the parity and the time-reversal operation refers to the following:
\\
\\
$\mathcal{P}: \hat{x} {\rightarrow}~ -\hat{x},~ \hat{p} ~{\rightarrow}~-\hat{p}$
\\
\\
$\mathcal{T}:\hat{x} ~{\rightarrow} ~\hat{x}, ~\hat{p}~ {\rightarrow}~-\hat{p},~{i} ~{\rightarrow}-{i}$
\\
\\
It should be noted that unlike hermitian Hamiltonians, $\mathcal{PT}$-symmetric Hamiltonians exhibit phase transition in the following sense. All the eigenvalues of the Hamiltonians described by Eq.~1 are real for ${\epsilon}\geq0$, a regime called unbroken $\mathcal{PT}$ symmetry. On the other hand, eigenvalues cease to be real, i.e. they become complex when, $-1<{\epsilon}<0$; this is known as the regime of broken $\mathcal{PT}$ symmetry. This is a common characteristic of parity-time symmetric Hamiltonians. There is a threshold value of the parameter, ${\epsilon}$, say, ${\epsilon}={\epsilon_{th}}$ at which the Hamiltonians enter from the broken to the unbroken $\mathcal{PT}$-regime or vice-versa. 
\section{\label{sec2} $\mathcal{PT}$-Optics: origin of the theoretical idea}

\captionsetup[wrapfigure]{textfont=small,justification=raggedright}
\begin{wrapfigure}{R}{0.15\textwidth}
\centering
\includegraphics[width=0.15\textwidth]{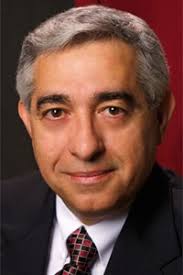}
{\caption*{Prof. Demetrios Christodoulides, CREOL}}
\end{wrapfigure}
In quantum mechanics, the so-called Schr\"{o}dinger equation (with $\hbar=m=1$) is given by $i \partial_t\psi=\hat{H}\psi$, where $\hat{H}=\hat{p}^2/2+V(\hat{x})$ and $\hat{p} {\rightarrow} - i\partial_x$. The Hamiltonian $H$ is parity-time symmetric if $[H,\mathcal{PT}]=0$, which basically means that we must have: $V(\hat{x})=V^{*}(-\hat{x})$. In other words,parity-time symmetry requires that the real part  ($R$) of such a potential is an even function of position $x$ whereas the imaginary part ($I$) is an odd function. Thus the Hamiltonian must have the form $\hat{H}=\hat{p}^2/2m+V_R(\hat{x})+i {\epsilon}V_I(\hat{x})$. Clearly if ${\epsilon=0}$, then the Hamiltonian is hermitian. It so happens that the Hamiltonian $H$ still exhibits real spectra as long as ${\epsilon}$ is below some threshold value, ${\epsilon_{th}}$, as discussed earlier, referring to the so-called unbroken $\mathcal{PT}$-regime. The physicists, in particular a group at CREOL, University of Central Florida, led by Professor Demetrios Chrostodoulides,was clever enough to observe the isomorphism between the Schr\"{o}dinger equation and the so-called paraxial equation of diffraction in optics,given by: $i\partial_z\phi+(1/2k) \partial_{xx}+k_0n \phi=0$, where $k_0=2\pi/\lambda$ and $k=k_0n_0$. $n=n_R(x)+in_I(x)$ is the complex refractive index. $n_R$ is the real refractive index profile and  $n_I$ represents the imaginary gain/loss profile. $\lambda$ is the wavelength of light in vacuum and $n_0$ represents the background refractive index. In general, $n_0\gg n_{R,I}$. It was quickly evident to the CREOL group that the optical system would be parity-time symmetric if the following conditions are satisfied: $n_R(x)=n_R(-x)$ and $n_I(x)=-n_I(-x)$. Subsequently, they came up with a proposal to connect $\mathcal{PT}$-symmetric quantum mechanics and optics \cite{2}. Obviously,  fulfilment of these conditions must have far reaching consequences! It turns out that one needs to synthesise artificial optical structures to fulfil the above mentioned conditions. Since, physically speaking, the imaginary part of the complex refractive index $n$, i.e. $n_I(x)$ refers to the loss (gain) profile of the structure at position, $x$ ,the condition $n_I(x)=-n_I(-x)$ demands that there must also have a corresponding gain (loss)  profile at position, $-x$, for the structure to be parity-time symmetric. In other words, $\mathcal{PT}$ symmetry is attained through a judicious inclusion of gain-loss dipoles in such structures. This issue will be more clear to the readers in the following section when we will discuss a parity-time symmetric directional coupler. When fabricated,these man-made engineered structures exhibit astonishing characteristics which could be utilised for various novel applications. The key point is that `loss' which is usually considered to be an `evil' in optical systems is no longer so, rather it turns out to be a blessing in this new kind of optics! Thus, nature demands that in non-Hermitian or parity-time symmetric optics, all the so-called basic three ingredients of optics, namely: refractive index, gain and loss are equally important. 
\section{\label{sec3} Understanding $\mathcal{PT}$-Optics via a directional coupler}
Many features of non-Hermitian optics could be understood with a directional coupler. A directional coupler is basically a pair of two parallel optical waveguides placed close to each other so that they are coupled to one another via the so-called evanescent coupling. Now the structure is parity time symmetric if one waveguide has loss, and then the other one must has equal gain or vice-versa (refer to Fig.~1). 

\captionsetup[figure]{labelfont=bf,textfont=small,justification=raggedright}
\begin{figure}
\includegraphics[height=3cm,width=3cm]{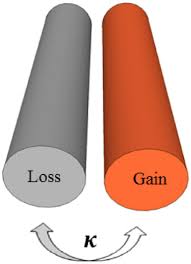}
\caption{\label{Fig1}Schematic of a parity-time symmetric directional coupler. The coupling co-efficient is denoted by $\kappa$}
\end{figure}
The system could be described by the following set of coupled differential equations:
\begin{align}
i\frac{da}{dt}-i g a+\kappa b=0, i\frac{db}{dt}+i g b+\kappa a=0
\end{align}
Here $\kappa$ and $g$ respectively are the coupling and the gain/loss parameter. On the other hand, $a$ and $b$ represents the optical fields in the gain and the loss waveguide respectively. It is easy to see that the Hamiltonian describing the system in $(a,b)$ basis is given by:
 \begin{equation}
H=\begin{pmatrix}ig&-\kappa\\
-\kappa&-ig\\
\end{pmatrix}
\end{equation} 

Clearly, the Hamiltonian $H$ in Eq.~3 is non-Hermitian. Also, it is parity-time symmetric, as it satisfies the commutation relation, $[H,\mathcal{PT}]=0$, where $\mathcal{P}=\begin{pmatrix} 0&1\\1&0\end{pmatrix}$ and $\mathcal{T}$ refers to taking the complex conjugation. The eigenvalues of the Hamiltonian is given by: $ E=\sqrt{\kappa^2-g^2}$. It means, as long as $\kappa>g$, the eigenvalues are real and the system is said to be in the unbroken $\mathcal{PT}$-regime. On the other hand, when $\kappa<g$, the eigenvalues become complex, i.e. the system enters into the so-called broken $\mathcal{PT}$-regime. The transition from the unbroken to the broken $\mathcal{PT}$-regime occurs at the critical value of the coupling parameter, $g_{th}=\kappa$ (please recall $\epsilon_{th}!$). This phase-transition point is also known as the exceptional point or the $\mathcal{PT}$-threshold. At the exceptional point, the eigenstates, also termed as {\it supermodes}, of the system coalesce. This is easy to see. The eigenstates of the Hamiltonian in Eq.~3 below the $\mathcal{PT}$-threshold (i.e. when $g<\kappa$) could easily be written as, with $sin{\Theta}=g/\kappa$, as follows:
\begin{equation}
\ket{1}_{below EP}=\begin{pmatrix}1\\e^{i\Theta}\end{pmatrix},
\ket{2}_{below EP}=\begin{pmatrix}1\\-e^{-i\Theta}\end{pmatrix}
\end{equation} 
On the other hand, above the $\mathcal{PT}$-threshold (i.e. when $g>\kappa$), the supermodes could be written as, with $cosh{\Theta}=g/\kappa$, as given below:
\begin{equation}
\ket{1}_{above EP}=\begin{pmatrix}1\\i e^{\Theta}\end{pmatrix},
\ket{2}_{above EP}=\begin{pmatrix}1\\i e^{-\Theta}\end{pmatrix}
\end{equation} 
Quite clearly, exactly at the exceptional point i.e. when $g=\kappa$, both the modes coalesce to  
$\ket{1,2}=\begin{pmatrix} 1\\i\end{pmatrix}$. In fact, the coalescence of eigenvectors at the exceptional point is a typical feature of all $\mathcal{PT}$-symmetric systems. This has immense physical implications in $\mathcal{PT}$-symmetric optical systems such as non-reciprocal behavior and power oscillations. 
\begin{figure}
\includegraphics[height=5cm,width=7cm]{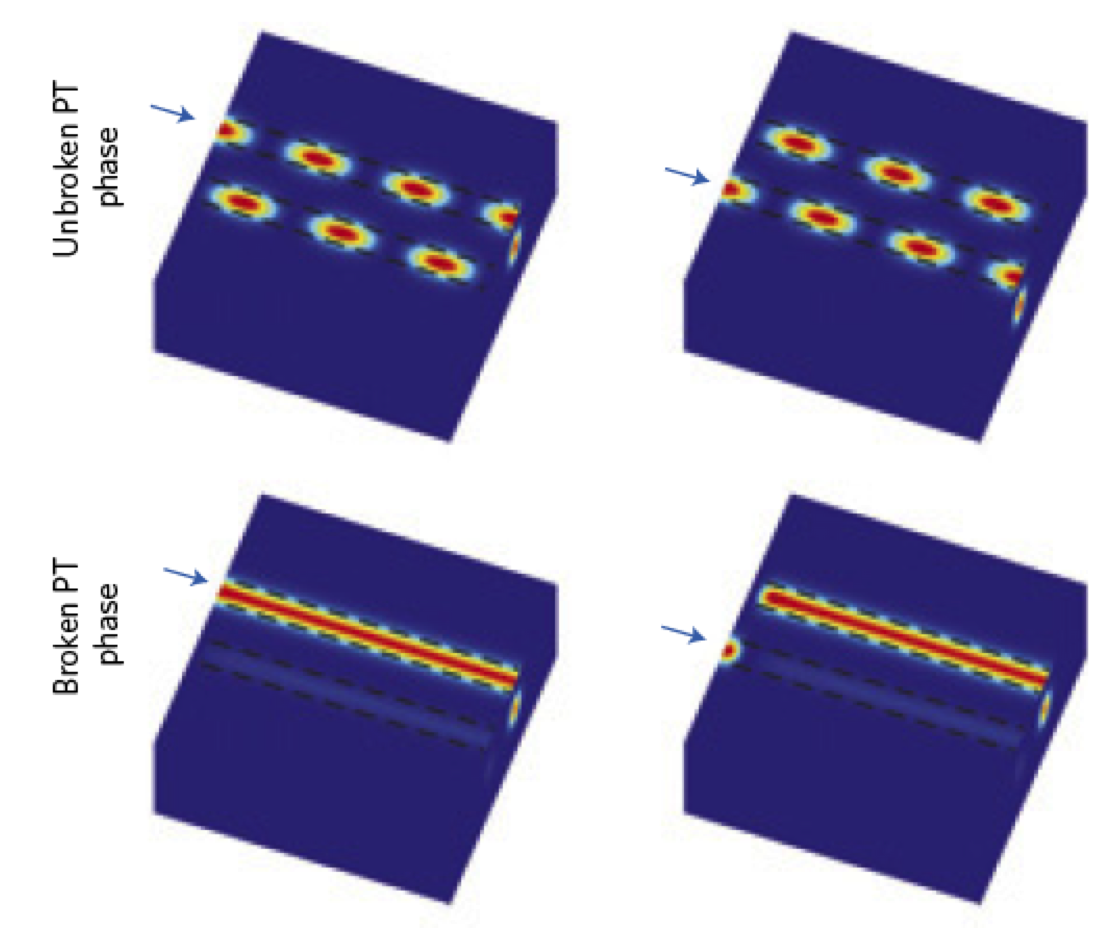}
\caption{\label{Fig2}Optical wave propagation when the system is excited at either channel 1 or channel 2. Light propagates in a non-reciprocal manner both below and above threshold. Adapted from ref.~\cite{4}.}
\end{figure}
Fig.~2 depicts the non-reciprocal nature of wave propagation in the $\mathcal{PT}$-coupler. Below the $\mathcal{PT}$-threshold, with exchange of the input channel from the gain to the loss waveguide, we obtain an entirely different output state. On the other hand, above the $\mathcal{PT}$-threshold light always leaves the coupler from the first waveguide, irrespective of the launch condition, albeit in a non-reciprocal manner. The physics behind this phenomenon could be traced back to the fact that above the $\mathcal{PT}$-threshold, eigenvalues of the Hamiltonian describing the system are complex, with the corresponding amplitudes either exponentially increasing or decaying, please refer to Eq.~5. Thus only one supermode effectively survives. The unexpected non-reciprocity behaviour of $\mathcal{PT}$-optical system is more clearly illustrated in Fig.~4. 
\begin{figure}
\includegraphics[height=2.5cm,width=6.5cm]{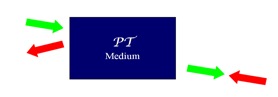}
\caption{\label{Fig3}Non-reciprocity in $\mathcal{PT}$synthetic structures}
\end{figure} 
Here the green arrows refer to a light beam that enters a synthetic $\mathcal{PT}$ structure from the left, and exits from the right. On the other hand, the red arrow on the right shows a light beam that enters the same $\mathcal{PT}$ system in exactly the opposite direction.  Astonishingly, the red beam fails to retrace the path of the green beam but instead emerges from the structure (left red arrow) along a different trajectory. This occurs owing to the {\it parity-time} invariance. Apart from the non-reciprocal behavior, one can observe that when the coupling of the waveguides is sufficiently strong, i.e. in the regime of unbroken $\mathcal{PT}$-symmetry, the system is in equilibrium and one observes Rabi oscillations (power oscillations), where the optical power oscillates back-and-forth between the two waveguides. On the other hand, when the coupling becomes too weak and the system enters into the broken $\mathcal{PT}$-regime, the Rabi oscillations cease and the system can no longer remain in equilibrium; the power then grows exponentially in one waveguide and decays exponentially in the other \cite{3}.
 \section{\label{sec4} Experimental Demonstrations of $\mathcal{PT}$-Optics and unusual features}
 The system of parity-time symmetric coupler discussed above was realized experimentally in the year 2010 by the CREOL group led by Christodoulides in collaboration with Detlef Kip's group in Germany and Segev's group in Israel, opening a new area in Optics. They have considered a $\mathcal{PT}$-symmetric coupled waveguide system based on Fe-doped LiNbO\textsubscript{3}. For the details of this remarkably  clever experiment, readers are referred to Ref.~\cite{4}. After this landmark experiment, plethora of activities took place both on theory and experimental fronts. Many more experimental demonstrations followed in various physical settings such as electronics, microwaves, mechanics, acoustics, atomic systems, optical mesh-lattice and so on \cite{5}. Some of the unusual features exhibited by $\mathcal{PT}$-optical systems are quite remarkable and noteworthy. Take for example, the phenomena of unidirectional invisibility \cite{6}! It has been theoretically demonstrated that parity-time symmetric periodic structures, near the $\mathcal{PT}$-threshold or the exceptional point, could act as a unidirectional invisible media. In this regime, the reflection from one end is diminished while it is enhanced from the other. This is illustrated in Fig.~4.The wave entering from the left channel does not recognize the existence of the periodic structure and goes through the sample entirely unaffected. On the other hand, a wave entering the same grating from the right, experiences enhanced reflection. Another commendable $\mathcal{PT}$-phase 
 \begin{figure}
\includegraphics[height=3cm,width=7cm]{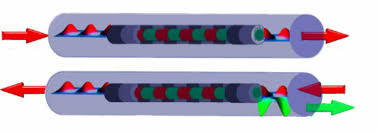}
\caption{\label{Fig4}Unidirectional invisibility of a $\mathcal{PT}$-symmetric grating.Adapted from ref.~\cite{6}.}
\end{figure}  
transition effect, which decisively shows that `loss' is not always an {\it evil}, is the loss-induced transparency \cite{7}! It is shown experimentally (please refer to Fig.~5) that an increase in {\it loss} enhances the overall transmission in a passive $\mathcal{PT}$-symmetric coupled waveguide channels, in the $\mathcal{PT}$-broken phase. 
  \begin{figure}
\includegraphics[height=4.5cm,width=7.5cm]{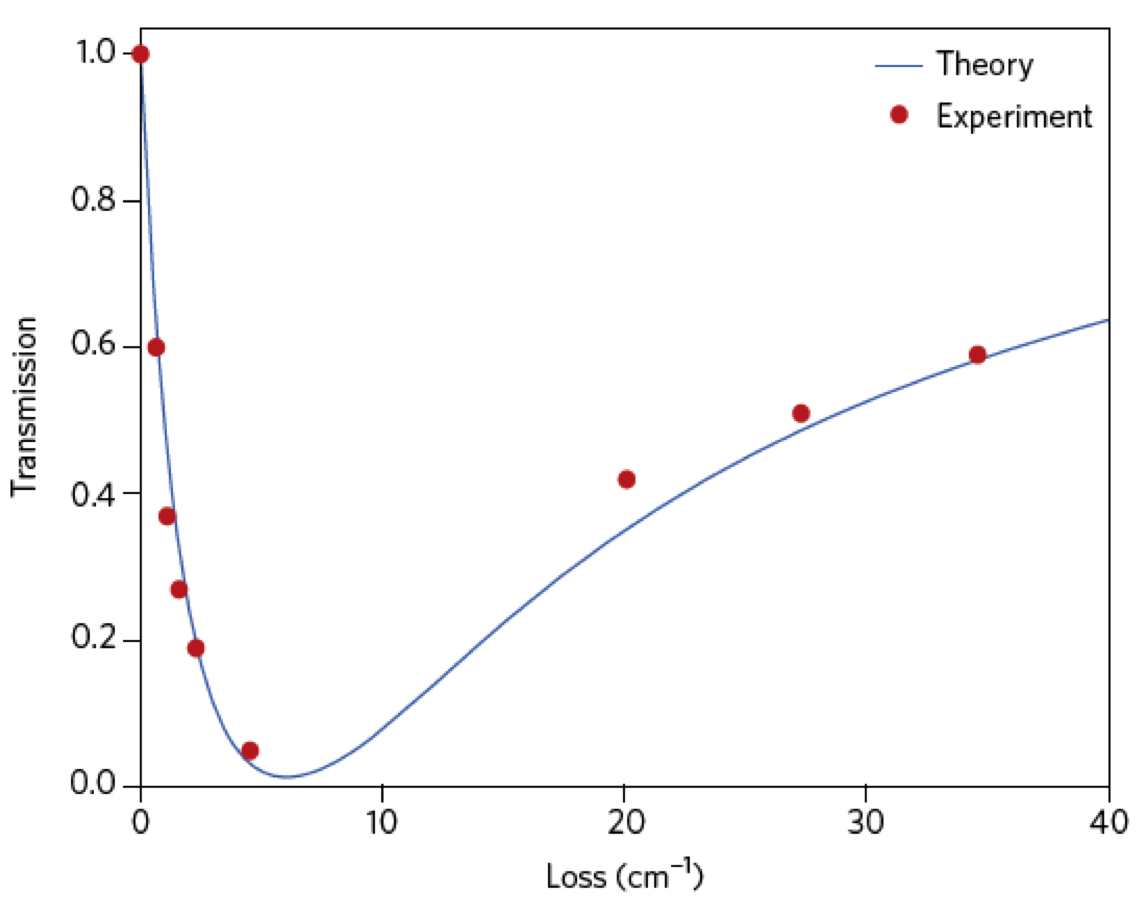}
\caption{\label{Fig5}Experimental observation of loss induced transparency in a passive $\mathcal{PT}$ complex system.Adapted from ref.~\cite{7}.}
\end{figure}  
What we have mentioned so far about the novelty or unusual characteristics of $\mathcal{PT}$-synthetic structure, in this short article, is extremely cursory.The area has developed quickly and come up with lots of remarkable promises. In particular, at the moment, it appears that the whole of Laser science and technology is getting revived \cite{8}. Lasers seem to provide the ultimate non-Hermitian playground where gain and loss are simultaneously engaged. The notion of $\mathcal{PT}$-symmetry as applied in optics, is no longer confined to optical systems only, it has now become an intense tool of explorations and studies even in condensed matter systems \cite{5}. It is quite clear that, considering the flurry of activities in $\mathcal{PT}$-symmetry, there will be many more breakthroughs in terms indelible commercial and practical applications in very near future.
 \setcounter{secnumdepth}{0}

\end{document}